\documentstyle[eqsecnum,aps]{revtex}
\newcommand{\be}{\begin{equation}}
\newcommand{\ee}{\end{equation}}
\newcommand{\bea}{\begin{eqnarray}}
\newcommand{\eea}{\end{eqnarray}}
\newcommand{\bml}{\begin{mathletters}}
\newcommand{\eml}{\end{mathletters}}
\newcommand{\pa}{\partial}

\newcommand{\bp}{\bar{\psi}}
\newcommand{\p}{\psi}
\newcommand{\bx}{\bar{\xi}}
\newcommand{\x}{\xi}

\newcommand{\dxy}{\delta(\vec{x}-\vec{y})}

\newcommand{\vx}{\vec{x}}
\newcommand{\vy}{\vec{y}}

\newcommand{\vk}{\vec{k}}

\newcommand{\g}{\gamma}

\newcommand{\e}{\epsilon}
\newcommand{\ve}{\varepsilon}

\begin{document}
\draft
\title{BFT Embedding of Second-Class Systems \cite{byline}}
\author{M. Fleck and H. O. Girotti}
\address{Instituto de F\'{\i}sica,
Universidade Federal do Rio Grande do Sul \\ Caixa Postal 15051, 91501-970  -
Porto Alegre, RS, Brazil.}

%\date{}
\maketitle
\begin{abstract}
The embedding procedure of Batalin, Fradkin, and Tyutin, which allows
to convert a second-class system into first-class, is pushed 
beyond the formal level. We study nonrelativistic as well as relativistic
systems. We explicitly construct, in all cases, the variables of the converted
first-class theory in terms of those of the corresponding second-class one. 
Moreover, we only conclude about the equivalence between these two different
kind of theories after compairing their respective spectra of excitations. 
\end{abstract}

\newpage

\section{Introduction}
\label{sec:level1}

Constrained systems possessing only second-class constraints, also known
as second-class systems, are to be quantized by abstracting the basic 
equal-time commutators (anticommutators) from the corresponding Dirac brackets,
the constraints thereby translating into strong operator relations\cite{Di1}. 
Then, the classical-quantum transition may be afflicted by ambiguities, which 
recognize as common origin the noncanonical structure of the Dirac brackets. 
This problem does not arise in connection with first-class 
systems, i.e., constrained systems whose constraints are all first-class, 
because in this case one can retain the canonical structure for the equal-time
commmutators (anticommutators) while imposing the constraints as restrictions 
on the states. Thus, it would be desirable to be able of converting any
second-class system into first-class. This is precisely what the Batalin-
Fradkin-Tyutin embedding procedure (BFT)\cite{BFT} does for us. 

However, the gauge invariant quantization procedure described above, which was
also proposed by Dirac\cite{Di1}, is not operational in most cases of 
physical interest. For instance, for non-Abelian gauge theories no one has yet
succeded in constructing a vacuum state of finite norm being anhilated by the 
corresponding Gauss law constraint\cite{Man}. In fact, the quantization of 
non-Abelian gauge field theories is to be performed in a fixed gauge, where
constraints and gauge conditions form a set of second-class constraints. Again,
the equal-time commutators are to be abstracted from the corresponding Dirac 
brackets\footnote{See, for instance, Refs.\cite{FVi1,Su,Gi1}}, which bring us 
back to the problem we were trying to avoid.

We believe that the importance of the BFT conversion mechanism rests more on 
the fact that it provides an efficient tool to generate a set of quantum 
mechanically equivalent theories. This paper is dedicated to a detailed study 
of this equivalence. 

We start, in Section 2, by proposing a second-class nonrelativistic system (a
toy model) whose quantization via Dirac brackets can be fully carried out and
is free of ambiguities. In Section 3, the BFT embedding 
procedure is used to generate the corresponding first-class counterpart. 
After verifying the existence of a unitary gauge, we implement a canonical 
transformation which enable us to construct all the phase-space variables of 
the first-class theory in terms of those of the second-class one. As we shall 
see, the first-class theory can be formulated in terms of gauge invariant 
variables only. Also, the number of degrees of freedom of the gauge invariant 
theory is larger than those of the second-class theory. Hence, it becomes non 
trivial to establish in what sense these two theories are classically and 
quantum mechanically equivalent. 
The outcomes from the St\"uckelberg embedding mechanism\cite{Stuc1}, for the 
same problem, are discussed in Section 4.

The self-dual (SD) model of Townsend, Pilch and Van Nieuwenhuizen\cite{TPN1} 
has recently served as a testing ground for appplying the
BFT embedding procedure in the relativistic case\cite{Kim,Rothe2,Rothe3}. 
After its conversion into first-class, the SD model appears to be quantum 
mechanically equivalent to the Maxwell-Chern-Simons (MCS)
theory\cite{Ha1,Jackiw2} when formulated in a Coulomb like gauge. In Section 5
we generalize the strategy of Section 3, based on canonical transformations,
and build up the phase-space variables of the MCS theory, in any arbitrary 
gauge, in terms of those of the SD model. Then, the SD and the MCS theories
will be shown to be rigorously equivalent irrespective of any gauge election.

The models in the previous sections only contain bosonic variables. In Section
6 we present the BFT embedding of the Proca-Wentzel field minimally coupled to
fermions. As known, this is a second-class theory possessing bosonic as well 
as fermionic second-class constraints. This time is a non-linear transformation
which enables us to write the converted first-class theory solely in terms of 
gauge invariant fields.   

The conclusions are contained in Section 7.

\section{A toy model}
\label{sec:level2}

We devote this section to study the classical and quantum dynamics of 
the nonrelativistic model whose Lagrangian is

\be
\label{201}
L \,=\,\frac{1}{2}\,m \omega\,\left( \,q^a\,\epsilon_{ab}\,{\dot{q}}^b\,
-\,\omega\,q^a\,g_{ab}\,q^b\,\right)\,\,\,.
\ee

\noindent
Here, $m$ is a mass parameter, $a$ runs from $1$ to $2$, $g_{ab}$ is the 
metric tensor of a two-dimensional Euclidean space and $\epsilon_{ab}$ is the 
completely antisymmetric tensor ($\epsilon_{12} = +1$). The first term in the 
right-hand side of (\ref{201}) is reminiscent of the Chern-Simons structure in
three space-time dimensions, while the second is a ``mass'' term.

The Lagrange equations of motion deriving from (\ref{201}),

\be
\label{202}
{\dot{q}}^a\,+\,\omega\,\epsilon^{ab}\,q_b\,=\,0 \,\,\,,
\ee

\noindent
tell us that the system under analysis is just a two dimensional particle in 
uniform circular motion, i.e.,

\be
\label{2021}
q^a\,q_a\,=\,\left(q^1\right)^2\,+\, \left(q^2\right)^2\,=\,constant\,\,\,,
\ee

\noindent
with (constant) angular velocity equal to $\omega$. Clearly, the energy 
($H^{(0)}$) and the angular momentum ($M^{(0)}$), 

\be
\label{203}
M^{(0)}\,=\,\frac{H^{(0)}}{\omega}\,=\,\frac{m\,\omega}{2}\,q^a \,q_a\,=\,
constant\,\,\,,
\ee

\noindent
are conserved quantities.

We shall denote by $p_a$ the momentum canonically conjugate to the coordinate 
$q^a$. Then, within the Hamiltonian framework, the system is characterized by
the primary constraints\footnote{Throughout this paper the sign of weak
equality ($\approx$) is used in the sense of Dirac\cite{Di1}.}

\be
\label{204}
T_a^{(0)}\,=\,p_a\,+\frac{1}{2}\,m\,\omega\,\epsilon_{ab}\,q^b\,
\approx\,0\,\,\,
\ee

\noindent
while the canonical Hamiltonian reads

\be
\label{205}
H^{(0)}\,=\,\frac{m \omega^2}{2}\,q^a\,q_a\,\,\,.
\ee

\noindent
Since the Poisson bracket ($[,]_P$)

\be
\label{206}
[ T_a^{(0)}\,,\,T_b^{(0)} ]_P\,=\,m\,\omega\,\epsilon_{ab}
\ee

\noindent
does not vanishes, the persistence in time of the primary contraints does not
give rise to secondary constraints and, hence, all the constraints are 
second-class. The usual counting reveals that only one independent degree of
freedom is present in (\ref{201}).  

The system can be quantized through the Dirac
bracket quantization procedure (DBQP)\cite{Di1}. According to this,
all phase space variables are to be promoted to operators obeying an equal-time
commutator algebra which is to be abstracted from the corresponding Dirac 
bracket algebra. For the model under analysis, one easily finds 
that\footnote{We shall not distinguish between a quantum operator and 
its classical counterpart.} 

\bml
\label{207}
\bea
&&[q^a\,,\,q^b]\,=\,-\,\frac{i\hbar}{m \omega}\,\epsilon^{ab}\,\,\,,
\label{mlett:a207}\\
&&[q^a\,,\,p_b]\,=\,\frac{i\hbar}{2}\,\delta^a_b\,\,\,,\label{mlett:b207}\\
&&[p_a\,,\,p_b]\,=\,-\frac{i\hbar \,m \omega}{4}\,\epsilon_{ab}\,\,\,.
\label{mlett:c207}
\eea
\eml

\noindent
As for the quantum mechanical Hamiltonian it can be read off directly from 
(\ref{205}), in view of the absence of ordering ambiguities in the 
classical-quantum transition. 

Within the algebra (\ref{207}) the constraints (\ref{204}) hold as strong
identities. Therefore, we use them to eliminate from the game $q^2$ and $p_2$.
We, also, define
 
\bml
\label{208}
\bea
&&q \,\equiv\,q^1\,=\,\frac{2}{m \omega}\,p_2\,\,\,,\label{mlett:a208}\\
&&p \,\equiv\,2\,p_1\,=\,-\,m \omega\,q^2\,\,\,,\label{mlett:b208}
\eea
\eml

\noindent
as the variables spanning the reduced phase-space ($\Gamma^{\ast}$)  
of the system. As required\cite{FVi1,Fa1}, they verify the canonical 
commutation relation $[q,p]=i\hbar$. Then, from (\ref{205})
follows that the reduced phase-space Hamiltonian is 

\be
\label{209}
H^{(0)\ast}\,=\,\frac{1}{2 m}\,p^{2}\,
+\,\frac{m}{2}\,\omega^2\,q^{2}\,\,\,,
\ee

\noindent
which describes an one dimensional harmonic oscillator with mass $m$ and 
proper frequency $\omega$. We shall designate by $|n>, n=0, +1, +2,...,$ the
eigenstates of $H^{(0)\ast}$, and by $E_n$, 

\be
\label{2091}
E_n = (n+1/2)\, \hbar |\omega|\,\,\,, 
\ee

\noindent
the corresponding eigenvalues.

The expression for the quantum mechanical angular momentum operator can not
be obtained by promoting (\ref{203}) to the quantum regime. Indeed, if one
insists on $M^{(0)\ast}\,=\,H^{(0)\ast}/\omega$, 
$M^{(0)\ast}$ fails to anhilate the ground state $|n=0>$. Because of this, we 
substitute (\ref{203}) by 

\be
\label{2092}
M^{(0)\ast}\,=\,\frac{H^{(0)\ast}}{\omega}\,-\,\frac{1}{2}\,\hbar |\omega| 
\,\,\,.
\ee

\noindent
To summarize, $M^{(0)\ast}$ and $H^{(0)ast}$ possess common eigenstates and the
eigenvalues of $M^{(0)}$ are $m_n \hbar$ with

\be
\label{2093}
m_n\,=\,n\,\e(\omega)\,\,\,,
\ee

\noindent
where $\e$ denotes the sign function.
 
We shall also need, for future purposes, the functional formulation of the 
quantum dynamics. The Green functions generating 
functional ($W$) for second-class systems was derived by Senjanovic\cite{Se}. 
Since in the present case the determinant 
$\det{[ T_a^{(0)}\,,\,T_b^{(0)} ]_P} $ is just a number, the expression for 
$W$ reduces to

\bea
\label{210}
W\,&=&\,{\cal N}\,\int [\prod_{a=1}^2 {\cal D}q^a] [\prod_{a=1}^2 {\cal D}p_a] 
\left(\prod_{a=1}^{2}\,\delta [p_a\,
+\,\frac{1}{2}\,m\,\omega\,\epsilon_{ab}\,q^b]\right)
\nonumber\\
& \times &\, \exp\left[\,i\,\int^{+\infty}_{-\infty}\,dt\,
\left( p_a {\dot q}^a\,- \,\frac{m \omega^2}{2}\,q^a q_a\right) \right]\,\,\,,
\eea

\noindent
where ${\cal N}$ is a normalization constant. After performing the momentum 
integrations one obtains

\be
\label{211}
W\,=\,{\cal N}\,\int [\prod_{a=1}^2 {\cal D}q^a] 
\, \exp\left[\,i\,\int^{+\infty}_{-\infty}\,dt\,m\,\omega\,
\left( q^a\,\epsilon_{ab}\, {\dot q}^b\,
- \,\omega\,q^a\,g_{ab}\, q^b\right) \right]\,\,\,,
\ee

\noindent
which sais that the effective Lagrangian coincides with (\ref{201}). A
further integration on $q^2$ yields

\be
\label{212}
W\,=\,{\cal N}\,\int\,[{\cal D}q] 
\, \exp \left[\,i\,\int^{+\infty}_{-\infty}\,dt\,
\left( \frac{m}{2}\,{\dot q}^{2}\,- \,\frac{m}{2}\,
\omega^2\,q^{2} \,\right) \right]\,\,\,,
\ee

\noindent
where we have used (\ref{mlett:a208}). This confirms that the reduced system is
an one dimensional harmonic oscillator. The analysis of the toy model as a 
second-class system is by now complete.

\section{BFT embedding of the toy model}
\label{sec:level3}

We next use the BFT\cite{BFT} procedure to convert the second-class system
described in the previous Section into first-class\footnote{For a
detailed description of the BFT procedure we refer the reader to the original
papers in Ref.\cite{BFT}.}. For this purpose, 
one starts by introducing an additional pair of canonical variables
(coordinate $u^a$ and momentum $s_a$) for each second-class constraint. The 
new constraints and the new Hamiltonian are found, afterwards, through an 
iterative scheme which, in the present case, ends after a finite number of 
steps. Presently, the BFT conversion procedure yields

\be
\label{301}
T_a^{(0)} \longrightarrow T_a \,=\,T_a^{(0)}\,+\,T_a^{(1)}\,=\,   
p_a\,+\,\epsilon_{ab}\,\left (\,\frac{1}{2}\,m\,\omega\,q^b\,
+\,\sqrt{m \omega}\,z^b\,\right)\,\approx\,0\,\,\,,
\ee

\be
\label{302}
H^{(0)} \longrightarrow H\,=\,H^{(0)}\,+\,H^{(1)}\,+\,H^{(2)}\,=\,
\frac{m \omega^2}{2}\,\left(\,q^a\,+\,\frac{1}{\sqrt{m \omega}}\,z^a\,
\right)^2\,\,\,,
\ee

\noindent
where the following definition

\be
\label{303}
z^a\,\equiv\,- \frac{1}{2}\,u^a\,-\,\epsilon^{ab}\,s_b\,\,\,,
\ee

\noindent
has been introduced. We emphasize that the $z^{a}$'s are composite 
objects whose Poissson bracket algebra,

\be
\label{3031}
\left[z^a\,,\,z^b\right]_P\,=\,-\,\e^{ab}\,\,\,,
\ee 

\noindent
derives from the canonical algebra

\bml
\label{3032}
\bea
&& \left[u^a\,,\,u^b\right]_P\,=\,0\,\,\,,\label{mlett:a3032}\\
&& \left[s_a\,,\,s_b\right]_P\,=\,0\,\,\,,\label{mlett:b3032}\\
&& \left[u^a\,,\,s_b\right]_P\,=\,\delta^a_b\,\,\,.\label{mlett:c3032}
\eea
\eml

\noindent
One can easily check that the new constraints and the new Hamiltonian are, as
required, strong under involution, i.e.,

\bml
\label{304}
\bea
&&[\,T_a\,,\,T_b\,]_P\,=\,0\,\,\,,\label{mlett:a304}\\
&&[\,T_a\,,\,H\,]_P\,=\,0\,\,\,.\label{mlett:b304}
\eea
\eml

\noindent
The converted system is, indeed, first-class and obeys an Abelian involution
algebra.

We construct next the unitarizing Hamiltonian ($H_U$) and the corresponding 
Green functions generating functional ($W_{\chi}$)\cite{BFT}. If we denote by

\be
\label{305}
\Psi\, \equiv \, {\bar{\cal C}}_a\,\chi^a\,-\,{\bar {\cal P}}_a\,\lambda^a
\,\,\,,
\ee

\be
\label{306}
\Omega\, \equiv {\bar {\pi}}_a\,{\cal P}^a\,+\,T_a\,{\cal C}^a\,\,\,,
\ee

\noindent
the gauge fixing fermion function and the BRST charge, respectively, one has 
that

\be
\label{307}
H_U\,=\,H\,-[\,\Psi\,,\,\Omega\,]_P\,\,\,.
\ee

\noindent
Here, ${\cal C}^a$ and ${\bar{\cal C}}_a$ are ghost coordinates 
and ${\bar{\cal P}}_a$ and ${\cal P}^a$ their respective canonical 
conjugate momenta. Furthermore, $\lambda^a$ is the Lagrange multiplier 
associated with the constraint $T_a$ and ${\bar{\pi}}_a$ is its 
canonical conjugate momentum. The gauge conditions $\chi^a$ are to be chosen 
such that 

\be
\label{308}
\det {[\, \chi^a\,,\,T_b\, ]_P}\,\neq\,0\,\,\,.
\ee

\noindent
The generating functional $W_{\chi}$, corresponding to the unitarizing 
Hamiltonian (\ref{307}), is

\be
\label{309}
W_{\chi}\,=\,{\cal N}\,\int\,[D\sigma]\,\exp(iA_U)\,\,\,,
\ee

\noindent
where the unitarizing action $A_U$ is given by

\be
\label{310}
A_U\,=\,\int^{+\infty}_{-\infty}\,dt\,
\left( p_a\,{\dot q}^a\,+\,s_a\,{\dot u}^a\, +\,{\bar {\pi}}_a\, 
{\dot {\lambda}}^a\,+\,{\bar {\cal C}}_a\,{\dot {\cal P}}^a
\,+\,{\bar {\cal P}}_a\,{\dot {\cal C}}^a\,-\,H_U \,\right)\,\,\,
\ee

\noindent
and the integration measure $[D\sigma]$ involves all the variables appearing in
$A_U$. To complete the characterization of the converted system, we mention
that under an infinitesimal supertransformation generated by $\Omega$, the 
phase space variables change as follows

\bml
\label{311}
\bea
\delta\,q^a\,& \equiv &\,[\,q^a\,,\,\Omega\,]_P\,\ve\,=\,{\cal C}^a\,\ve
\,\,\,,\label{mlett:a311}\\ 
\delta\,p_a\,& \equiv &\,[\,p_a\,,\,\Omega\,]_P\,\ve\,=\,\frac{m \omega}{2}\,
\epsilon_{ab}\,{\cal C}^b\,\ve\,\,\,,\label{mlett:b311}\\
\delta\,u^a\,& \equiv &\,[\,u^a\,,\,\Omega\,]_P\,\ve\,=\,\sqrt{m \omega}\,
{\cal C}^a\,\ve\,\,\,,\label{mlett:c311}\\
\delta\,s_a\,& \equiv &\,[\,s_a\,,\,\Omega\,]_P\,\ve\,
=\,-\,\sqrt{\frac{m \omega}{4}}\,\epsilon_{ab}\,
{\cal C}^b\,\ve\,\,\,.\label{mlett:d311}\\
\delta\,\lambda^a\,& \equiv &\,[\,\lambda^a\,,\,\Omega\,]_P\,\ve\,
=\,{\cal P}^a\,\ve \,\,\,,\label{mlett:e311}\\
\delta\,{\bar {\cal C}}_a\,& \equiv &\,[\,{\bar {\cal C}}_a\,,\,\Omega\,]_P
\,\ve\,=\,{\bar {\pi}}_a\,\ve\,\,\,,\label{mlett:f311}\\
\delta\,{\bar {\cal P}}_a\,& \equiv &\,[\,{\bar {\cal P}}_a\,,\,\Omega\,]_P
\,\ve\,=\,T_a\,\ve\,\,\,,\label{mlett:g311}\\
\delta {\bar {\pi}}_a\,& \equiv & \,[\,{\bar {\pi}}_a\,,\,\Omega\,]_P\,
\ve \,=\,0 \,\,\,,\label{mlett:h311}\\
\delta {\cal C}^a\,& \equiv &\,[\,{\cal C}^a\,,\,\Omega\,]_P\,\ve \,=\,0
\,\,\,,\label{mlett:i311}\\
\delta {\cal P}^a\,& \equiv &\,[\,{\cal P}^a\,,\,\Omega\,]_P\,\ve \,=\,0
\,\,\,,\label{mlett:j311}
\eea
\eml

\noindent
where $\ve$ is an infinitesimal fermionic parameter.

We now focus on Eq.(\ref{309}) and restrict ourselves to consider gauge 
conditions which do not depend upon $\lambda^a$ and/or ${\bar {\pi}}_a$. 
Then, the rescaling $\chi^a \rightarrow \chi^a/\beta$, 
${\bar {\pi}}_a \rightarrow \beta {\bar {\pi}}_a$ and 
${\bar {\cal C}}_a \rightarrow \beta {\bar {\cal C}}_a$ allows, at 
the limit $\beta \rightarrow 0$, to carry out all the integrals
over the ghosts and multiplier variables\cite{FVi2,H1}, with the result

\bea
\label{312}
W_{\chi}\,&=&\,{\cal N}\,\int [\prod_{a=1}^2 {\cal D}q^a] \,
[\prod_{a=1}^2 {\cal D}p_a]\, [\prod_{a=1}^2 {\cal D}u^a]\,
[\prod_{a=1}^2 {\cal D}s_a]\,\det {[\, \chi^a\,,\,T_b\, ]_P}
\nonumber\\ 
& \times &\,
\left(\prod_{a=1}^{2}\,\delta [\,T_a\,]\right)
\left(\prod_{a=1}^{2}\,\delta [\,\chi^a\,]\right)
 \exp\left[\,i\,\int^{+\infty}_{-\infty}\,dt\,
\left( p_a {\dot q}^a\,+ \,s_a {\dot u}^a\,-\,H\,\right)\right]\,\,\,.
\eea
 
\noindent
The structure of the Hamiltonian (see Eq.(\ref{302})) suggests the
change of variables
$u^a \rightarrow u^{\prime a} = z^a$, $s_a \rightarrow s^{\prime}_{a} = s_a$,
whose jacobian is a nonvanishing real number. Since $H$ does not depend upon
$s^{\prime}_{a}$, the corresponding integration is straightforward and after
performing it one obtains

\bea
\label{313}
W_{\chi}\,&=&\,{\cal N}\,\int [\prod_{a=1}^2 {\cal D}q^a] \,
[\prod_{a=1}^2 {\cal D}p_a]\,[\prod_{a=1}^2 {\cal D}z^a]\,
\det {[\, \chi^a\,,\,T_b\, ]_P} \nonumber\\ 
& \times &\,
\left(\prod_{a=1}^{2}\,\delta [\,T_a\,]\right)
\left(\prod_{a=1}^{2}\,\delta [\,\chi^a\,]\right)
 \exp\left[\,i\,\int^{+\infty}_{-\infty}\,dt\,
\left( p_a {\dot q}^a\,+ \frac{1}{2}\,z^a\,\epsilon_{ab}\,{\dot z}^b\,
-\,H\,\right)\right]\,\,\,.
\eea

\noindent
Notice that, up to a surface term, $1/2 \int dt 
z^a \epsilon_{ab} {\dot z}^b$ can be rewritten in the standard canonical form 
$\int dt z^1 {\dot z}^2$. The interpretation of $z^1$ as the canonical 
conjugate momentum of $z^2$, suggested by (\ref{3031}), is then possible but
not mandatory. Finally, we take advantage of the functions $\delta [T_a]$ to 
carry out the integrals on $p_a$, thus arriving at the following final 
expression for $W_{\chi}$

\be
\label{314}
W_{\chi}\,=\,{\cal N}\,\int [\prod_{a=1}^2 {\cal D}q^a] \,
[\prod_{a=1}^2 {\cal D}z^a]\,\det {[\, \chi^a\,,\,T_b\, ]_P}  
\left(\prod_{a=1}^{2}\,\delta [\,\chi^a\,]\right)
 \exp\left(i \int^{+\infty}_{-\infty}\,dt\,L_{BFT} \right)\,\,\,,
\ee

\noindent
where

\be
\label{315}
L_{BFT}\,=\,\frac{m \omega}{2} \, 
\left(\,x^a \, \epsilon_{ab} \,{\dot {x}}^b\,-\,
\omega\,\,x^a \,g_{ab}\,\,x^b \,\right)\,\,\,
\ee

\noindent
and 

\be
\label{316}
x^a \,\equiv\, q^a + \frac{1}{\sqrt{m \omega}} z^a\,\,\,.
\ee

\noindent
One can easily verify, from (\ref{303}), (\ref{mlett:a311}), (\ref{mlett:c311})
and (\ref{mlett:d311}), that $x^a$ is gauge invariant.

Clearly, the subsidiary conditions $z^a \approx 0$ enables one to recover the
original second-class theory (see (\ref{210})) and, therefore, define the 
unitary gauge. This is enough to secure that the first-class theory (\ref{314})
is equivalent to the second-class theory (\ref{210}) from which we started. 

Most investigations on the BFT embedding procedure end at the level of
our expression (\ref{313}). This leave out of consideration a whole of
possibilities whose analysis is one of our purposes in this work. To exemplify
what we mean by this, we perform in (\ref{312}) the canonical transformation 

\bml
\label{317}
\bea
q^a\,& \rightarrow & \, Q^a \, \equiv \, \frac{1}{2}\,q^a\,+\,
\frac{1}{m \omega}\,\e^{ab}\,p_b \,-\,\frac{1}{\sqrt{4 m \omega}}\,u^a\,+
\,\frac{1}{\sqrt{m \omega}}\, \e^{ab}\,s_b \,\,\,,\label{mlett:a317} \\  
p_a\,& \rightarrow & \, P_a \, \equiv \, \frac{1}{2}\,p_a\,-\,
\frac{m \omega}{4}\,\e^{ab}\,q^b \,
-\,\sqrt{\frac{m \omega}{16}}\,\e_{ab}\, u^b\,
- \,\sqrt{\frac{m \omega}{4}}\,s_a \,\,\,,\label{mlett:b317} \\  
u^a\,& \rightarrow & \, U^a \, \equiv \, \frac{1}{2}\,q^a\,-\,
\frac{1}{m \omega}\,\e^{ab}\,p_b \,+\,\frac{1}{\sqrt{4 m \omega}}\,u^a\,+
\,\frac{1}{\sqrt{m \omega}}\, \e^{ab}\,s_b \,\,\,,\label{mlett:c317} \\  
s_a\,& \rightarrow & \, S_a \, \equiv \,\frac{1}{2}\,T_a 
\,\,\,,\label{mlett:d317}
\eea
\eml

\noindent
which, after some algebra, enables one to rewrite

\bea
\label{318}
W_{\chi}\,&=&\,{\cal N}\,\int [\prod_{a=1}^2 {\cal D}Q^a] \,
[\prod_{a=1}^2 {\cal D}P_a]\, [\prod_{a=1}^2 {\cal D}U^a]\,
[\prod_{a=1}^2 {\cal D}S_a]\,\det {[\, \chi^a\,,\,S_b\, ]_P}
\nonumber\\ 
& \times &\,
\left(\prod_{a=1}^{2}\,\delta [\,S_a\,]\right)
\left(\prod_{a=1}^{2}\,\delta [\,\chi^a\,]\right)
 \exp\left[\,i\,\int^{+\infty}_{-\infty}\,dt\,
\left( P_a {\dot Q}^a\,+ \,S_a {\dot U}^a\,-\,K\,\right)\right]\,\,\,,
\eea

\noindent
where the transformed Hamiltonian $K$ is given by

\be
\label{319}
K\,=\,\frac{1}{2 m}\,P_a\,g^{ab}\,P_b\,+\,\frac{\omega}{2}\,Q^a\,\e_a^b\,P_b\,
+\,\frac{m \omega^2}{8}\,Q^a\,g_{ab}\,Q^b\,\,\,.
\ee

\noindent
Notice that $Q^a$, $P_a$, and $S_a$ are gauge invariant phase-space variables,
as can be seen from (\ref{311}) and (\ref{317}). Moreover, the canonical 
transformation (\ref{317}) has been chosen in such a way that the constraints 
are simply given by the equations $S_a = 0$. Thus, only the coordinates 
$U^a$ are affected by gauge transformations and, furthermore, $U^a \approx 0$ 
are admissible gauge conditions, to be selected from now on. Hence, after 
carrying out the $S$,$U$ and $P$ integrals in (\ref{319}) one arrives at

\be
\label{320}
W_{U=0}\,=\,{\cal N}\,\int\, [\prod_{a=1}^2 {\cal D}Q^a] \,\exp \left(\,i\,
\int_{-\infty}^{+\infty}\,dt\,L_{U=0}\right)\,\,\,,
\ee

\noindent
where the effective Lagrangian ($L_{U=0}$)

\be
\label{321}
L_{U=0}\,=\,\frac{m}{2}\,{\dot Q}^a\,g_{ab}\,{\dot Q}^b\,
-\,\frac{m\,\omega}{2}\,Q^a\,\e_{ab}\,{\dot Q}^b\,\,\,,
\ee

\noindent
only contains gauge invariant degrees of freedom.

Observe that the ``mass'' term in (\ref{201}) has been replaced by a
standard kinetic energy term. Then, $L_{U=0}$ does not describe a constrained
system but a regular one possessing truly two independent degrees of 
freedom. On the other hand, $L$ involves, as we already pointed out, only one
independent degree of freedom. Since the BFT conversion procedure should no
alter the physics, $L_{U=0}$ and $L$ must be equivalent. We shall
next investigate this equivalence in detail. 

Let us first look at the classical dynamics arising from $L_{U=0}$. The
Lagrange equations of motion are found to read

\be
\label{322}
{\ddot Q}^a\,+\,\omega\,\e^{ab}\,{\dot Q}_b\,=\,0\,\,\,,
\ee

\noindent
implying that

\be
\label{323}
{\dot Q}^a\,+\,\omega\,\e^{ab}\,Q_b\,=\,C^a\,\,\,,
\ee

\noindent
where
$C^a, a=1,2$ are constants of motion. The Cartesian form of the trajectory is
easily found to be

\be
\label{324}
\left(\,Q^1\,+\,\frac{C_2}{\omega}\,\right)^2\,+\,
\left(\,Q^2\,-\,\frac{C_1}{\omega}\,\right)^2\,=\,constant\,\,\,,
\ee

\noindent
which are just circles centered at $Q^1 = - C^2, Q^2 = C^1$. Also, the energy 
($K$) and the angular momentum ($M$),

\be
\label{325}
K\,=\,\frac{m}{2}\,{\dot Q}^a\,{\dot Q}_a\,\,\,,
\ee

\be
\label{326}
M\,=\,\frac{E}{\omega}\,-\,\frac{m}{2 \omega}\,C^a C_a\,\,\,,
\ee

\noindent
turn out to be conserved quantities. By comparing these results with the
corresponding ones in Section 2, we conclude that the second-class rotator
was converted into a first-class system just by turning arbitrary the position
of the center of rotation.

We turn next into quantizing $L_{U=0}$. For a regular (unconstrained) system
the equal-time commutation algebra must be abstracted from the corresponding
Poisson bracket algebra. Then,

\bml
\label{3261}
\bea
&&\left[\,Q^a\,,\,Q^b\right]\,=\,0\,\,\,,\label{mlett:a3261}\\ 
&&\left[\,Q^a\,,\,P_b\right]\,=\,i \hbar\,\delta^a_b
\,\,\,,\label{mlett:b3261}\\ 
&&\left[\,P_a\,,\,P_b\right]\,=\,0\,\,\,.\label{mlett:c3261}
\eea
\eml 

\noindent
Due to the absence of ordering ambiguities, the Hamiltonian operator can be
read off from (\ref{319}). On the other hand, 
in terms of phase-space variables the angular momentum operator is found to 
read

\be
\label{3262}
M\,=\,Q_a\,\e^{ab}\,P_b\,\,\,.
\ee

It will prove convenient to introduce destruction ($A_{\pm}$) and creation 
($A^{\dag}_{\pm}$) operators of definite helicity defined as

\bml
\label{327}
\bea
&&A_{\pm}\,\equiv\,\frac{1}{\sqrt {2}}\,\left(\,A_1\,\mp\,i A_2\,\right)\,\,\,,
\label{mlett:a327}\\
&&A^{\dag}_{\pm}\,\equiv\,\frac{1}{\sqrt {2}}\,
\left(\,A^{\dag}_1\,\pm\,i A^{\dag}_2\,\right)\,\,\,, \label{mlett:b327}
\eea
\eml 

\noindent
where $A_a$ and $A^{\dag}_a$, $a=1,2$, are, respectively, standard 
destruction and creation operators, i.e.,

\bml
\label{328}
\bea
&&A_a\,\equiv\, \left( \frac{m |\omega|}{4 \hbar}\right)^{\frac{1}{2}}\,Q_a\,
+\,i\,\left( m \hbar |\omega|\right)^{-\frac{1}{2}}\,P_a\,\,\,,
\label{mlett:a328}\\
&&A^{\dag}_a\,\equiv\, 
\left( \frac{m |\omega|}{4 \hbar}\right)^{\frac{1}{2}}\,Q_a\,
-\,i\,\left( m \hbar |\omega|\right)^{-\frac{1}{2}}\,P_a\,\,\,.
\label{mlett:b328} 
\eea
\eml

\noindent
The equal-time algebra verified by $A_{\pm}$ and $A^{\dag}_{\pm}$,
 
\bml
\label{329}
\bea
&&\left[\,A_r\,,\,A_s\,\right]\,=\,0\,\,\,,\label{mlett:a329}\\
&&\left[\,A^{\dag}_r\,,\,A^{\dag}_s\,\right]\,=\,0\,\,\,,\label{mlett:b329}\\
&&\left[\,A_r\,,\,A^{\dag}_s\,\right]\,=\,\delta_{rs}\,\,\,,\label{mlett:c329}
\eea
\eml
 
\noindent
where $r = +,-$ and $s = +,-$, follows from (\ref{3261}), (\ref{327}) and 
(\ref{328}).

Now, the Hamiltonian and the angular momentum operators can be cast as

\be
\label{330}
K\,=\,\left(\,N_{+}\,+\,N_{-} \,\right) \frac{\hbar |\omega|}{2}\,+\,
\frac{\hbar |\omega|}{2}\,+\,\left(\,N_{+}\,-\,N_{-}\,\right)
\frac{\hbar \omega}{2}\,\,\,,
\ee

\be
\label{331}
M\,=\,\hbar\, \left(\,N_{+}\,-\,N_{-}\,\right)\,\,\,,
\ee

\noindent
where

\bml
\label{332}
\bea
&&N_{+}\,\equiv\,A^{\dag}_{+} A_{+}\,\,\,,\label{mlett:a332}\\
&&N_{-}\,\equiv\,A^{\dag}_{-} A_{-}\,\,\,.\label{mlett:b332}
\eea
\eml

\noindent
We shall denote by $|n_{+} n_{-}>$ the common eigenstates of the hermitean 
commuting operators $N_{+}$ and $N_{-}$. They are labeled by the semidefinite
positive integers $n_{+},n_{-}$. Then, the eigenvalue problems for
$K$ and $M$ read, respectively,

\bml
\label{333}
\bea
K \, |n_{+} n_{-}>\,=\,E_{n_{+},n_{-}}\,|n_{+} n_{-}>\,\,\,,
\label{mlett:a333}\\
M \, |n_{+} n_{-}>\,=\,m_{n_{+},n_{-}}\,\hbar\,|n_{+} n_{-}>\,\,\,,
\label{mlett:b333}
\eea
\eml

\noindent
where

\be
\label{334}
E_{n_{+},n_{-}}\,=\,\left\{ \frac{\left[\,1\,+\,\e(\omega)\,\right]}{2}
\,n_{+}\,+\,\frac{\left[\,1\,-\,\e(\omega)\,\right]}{2}\,n_{-}\,
+\,\frac{1}{2}\right\}\,\hbar |\omega|\,\,\,,
\ee

\noindent
and

\be
\label{335}
m_{n_{+},n_{-}}\,=\,n_{+}\,-\,n_{-}\,\,\,.
\ee

\noindent
The dependence of $E_{n_{+},n_{-}}$ on $n_{+}$ and $n_{-}$ is rather peculiar.
In fact, for $\omega > 0$ the right hand side of Eq.(\ref{334}) reduces to 
$(n_{+} + 1/2)\hbar|\omega|$, while for $\omega < 0$ it goes into 
$(n_{-} + 1/2)\hbar|\omega|$. Hence, in either case, the energy eigenvalue 
spectrum is that of the second-class system (see Eq.(\ref{2091})). As for the
angular momentum, we observe that $m_{n_{+},n_{-}}$ depends simultaneously on 
$n_{+}$ and $n_{-}$, irrespective of the sign of $\omega$. Thus, the energy 
levels of the first-class system are (infinitely) degenerate, whereas those of
the second-class system are not (see Eq.(\ref{2093})). For both systems, the
range of eigenvalues of the angular momentum operator is the same. It is
in this sense that these systems are quantum mechanically equivalent. 

Before closing this Section, we would like to confirm that the degeneracy
of the energy levels is related to the arbitrariness in the position of the 
point around which the classical motion takes place. To this end, we start by
noticing that the right hand sides in Eqs.(\ref{319}) and (\ref{3262}) can 
both be diagonalized by means of the canonical transformation

\bml
\label{336}
\bea
Q^1 \,& = & \,\frac{1}{\sqrt 2}\,(\,\eta\,
+\,{\sqrt 2}\,\rho\,)\,\,\,,\label{mlett:a336}\\
Q^2 \,& = & \,\frac{{\sqrt 2}}{m \omega}\,
(\,p_{\eta}\,-\,\frac{1}{{\sqrt 2}}\,p_{\rho}\,)\,\,\,,\label{mlett:b336}\\
P_1 \,& = & \,\frac{1}{\sqrt 2}\,
(\,p_{\eta}\,+\,\frac{1}{{\sqrt 2}}\,p_{\rho}\,)\,\,\,,\label{mlett:c336}\\
P_2 \,& = &-\,\frac{m \omega}{2\,\sqrt 2}\,
(\,\eta\,-\,{\sqrt 2}\,\rho\,)\,\,\,.\label{mlett:d336}
\eea
\eml

\noindent
The replacement of (\ref{336}) into (\ref{319}) and (\ref{3262}) yields,
respectively, 

\be
\label{337}
K\,=\,\frac{1}{2 m}\,{p_{\rho}}^2\,+\,\frac{m}{2}\, \omega^2\,{\rho}^2\,\,\,,
\ee

\noindent
and

\be
\label{338}
M\,=\,\frac{1}{\omega}\,K\,-\,\frac{1}{\frac{\omega}{2}}
\left[\frac{{p_{\eta}}^2}{2 m}\,+\,\frac{m}{2}\left(
{\frac{\omega}{2}}\right)^2\,\eta^2\,\right]\,\,\,.
\ee 

\noindent
The decoupling of the sector $\eta$, $p_{\eta}$ from the dynamics should be
noticed. Moreover, from (\ref{326}) and (\ref{338}) follows that
the constants of motion $C^a, a=1,2$ can be written in terms of $\eta$ and
$p_{\eta}$ as 

\bml
\label{3381}
\bea
&&C^1\,=\,\frac{{\sqrt 2}}{m}\,p_{\eta}\,\,\,,\label{mlett:a3381}\\
&&C^2\,=\,-\,\frac{\omega}{{\sqrt 2}}\,\eta\,\,\,.\label{mlett:b3381}
\eea
\eml

\noindent
In fact, $\eta$ and $p_{\eta}$ are, up to proportionality constants, the
Noether charges associated with the invariance of $L_{U=0}$ under the global
translations $Q^a \rightarrow Q^a+C^a$. As already pictured, what we have
at hand is a rotator whose location in the plane $Q^1$ and $Q^2$ is arbitrary.

This new formulation of the converted model is to be quantized, again, by 
abstracting the
equal-time commutators from the corresponding Poisson brackets. The only
nonvanishing commutators, then, are

\bml
\label{339}
\bea
&&\left[\,\rho\,,\,p_{\rho}\,\right]\,=\,i \hbar\,\,\,,\label{mlett:a339}\\
&&\left[\,\eta\,,\,p_{\eta}\,\right]\,=\,i \hbar\,\,\,,\label{mlett:b339}
\eea
\eml

\noindent
and, consequently, from (\ref{337}), (\ref{338}) and (\ref{339}) one obtains 

\be
\label{340}
\left[\,K\,,\,M\,\right]\,=\,\left[\,K\,,\,\eta\,\right]\,
=\,\left[\,K\,,\,p_{\eta}\,\right]\,=\,0\,\,\,.
\ee

\noindent
Thus, we have four conserved observables ($K$, $M$, $\eta$, and $p_{\eta}$)
but not all of them are mutually commuting, since

\bml
\label{341}
\bea
&&\left[\,M\,,\,\eta\,\right]\,=\,\frac{2 i \hbar}{m
\omega}\,p_{\eta}\,\neq\,0 \,\,\,,\label{mlett:a341}\\
&&\left[\,M\,,\,p_{\eta}\,\right]\,=\,-\,\frac{i \hbar m \omega}{2}
\,\eta\,\neq\,0 \,\,\,.\label{mlett:b341}
\eea
\eml

\noindent
We select $K$ and $M$ as the maximal set of commuting observables. Their
common eigenstates will be labeled by the corresponding eigenvalues,
$E_n$ and $m_{n,{\bar n}}$, respectively. According to (\ref{337}) the
Hamiltonian $K$ is that of an harmonic oscillator of mass $m$ and proper
frequency $|\omega|$. Therefore,

\be
\label{342}
E_n\,=\,\left(\,n\,+\,\frac{1}{2}\,\right)\,\hbar |\omega|\,\,\,,
\ee

\noindent
where $n$ is a positive semidefinite integer. This confirms that the energy
eigenvalue spectrum is that of the second-class system (see (\ref{2091})). As
for the angular momentum, we observe that the first term in the right-hand side
of (\ref{338}) is just $K$ whereas the second is, up to a sign, proportional to
the Hamiltonian operator of an harmonic oscillator of mass $m$ and proper
frequency $|\omega|/2$. Hence,

\be
\label{343}
m_{n,{\bar n}}\,=\,(\,n\,-\,{\bar n}\,)\,\e(\omega)\,\,\,,
\ee

\noindent
where $n$ and ${\bar n}$ are positive semidefinite integers. As it must be the
case, this is in agreement with (\ref{335}). It is clear that the degeneracy of
the energy levels $E_n$ is due to the presence of ${\bar n}$ in (\ref{343}).
In turn, ${\bar n}$ originates from the harmonic oscillator in the sector 
$\eta$ and $p_{\eta}$. As we already pointed out, these are the variables that,
in the classical limit ($\hbar \rightarrow 0$), determine the location
of the center of rotation in the plane $Q^1$, $Q^2$.

\section{St\"uckelberg embedding of the discrete system}
\label{sec:level4}

The aim of this Section is to compare the outcomes of the BFT and the
St\"uckelberg\cite{Stuc1} embeddings in connection with the nonrelativistic
system defined by (\ref{201}). The St\"uckelberg embedding consists in
replacing, in (\ref{201}), $q^a \rightarrow q^a+y^a$, thus obtaining

\be
\label{401}
L_S \,=\,\frac{1}{2}\,m \omega\,\left[(\,q^a\,+\,y^a\,)\,\epsilon_{ab}\,
(\,{\dot{q}}^b\,+\,{\dot{y}}^b\,)\,-\,\omega\,
(\,q^a\,+\,y^a\,)\,g_{ab}\,(\,q^b\,+\,y^b\,)\right]\,\,\,. 
\ee

\noindent
We shall designate by $p_a$ and $w_a$ the momenta canonically conjugate to
$q^a$ and $y_a$, respectively. From the defining equations for $p_a$ and $w_a$
follows that the extended system is characterized by the primary second-class
constraints 

\be
\label{402}
\Omega_a\,=\,w_a\,
+\,\frac{1}{2}\,m\,\omega\,\e_{ab}\,(\,q^b\,+\,y^b\,)\,\approx\,0\,\,\,,
\ee

\noindent
the primary first-class constraints

\be
\label{403}
{\bar T}_a\,=\,p_a\,-\,w_a\,\approx\,0\,\,\,,
\ee

\noindent
and the canonical Hamiltonian

\be
\label{404}
H_S\,= \,\frac{m \omega^2}{2}\,(\,q^a\,+\,y^a)\,(\,q_a\,+\,y_a)\,\,\,.
\ee

\noindent
There are no secondary constraints. From (\ref{403}) follows 
that the infinitesimal gauge transformations generated by the first-class
constraints leave the Lagrangian (\ref{401})invariant, as must be the case. 
The second-class constraints can be eliminated by introducing partial Dirac 
brackets ($\Delta$-brackets) with respect to them\cite{Di1,FVi1,Su,Gi1}. 
For the nonvanishing $\Delta$-brackets one obtains

\bml
\label{405}
\bea
\left[\,q^a\,,\,p_b\,\right]_{\Delta}\,&=&\,\delta^a_b\,\,\,,\label{mlett:a405}\\ 
\left[\,p_a\,,\,p_b\,\right]_{\Delta}\,&=&\,- \,\frac{m \omega}{4}\,\e_{ab}
\,\,\,,\label{mlett:b405}\\ 
\left[\,p_a\,,\,y^b\,\right]_{\Delta}\,&=&\,\frac{1}{2}\,\delta^a_b
\,\,\,,\label{mlett:c405}\\ 
\left[\,p_a\,,\,w_b\,\right]_{\Delta}\,&=&\,- \,\frac{m \omega}{4}\,\e_{ab}
\,\,\,,\label{mlett:d405}\\ 
\left[\,y^a\,,\,y^b\,\right]_{\Delta}\,&=&\,- \,\frac{1}{m \omega}\,\e^{ab}
\,\,\,,\label{mlett:e405}\\ 
\left[\,y^a\,,\,w_b\,\right]_{\Delta}\,&=&\,\frac{1}{2}\,\delta^a_b
\,\,\,,\label{mlett:f405}\\ 
\left[\,w_a\,,\,w_b\,\right]_{\Delta}\,&=&\,- \,\frac{m \omega}{4}\,\e_{ab}
\,\,\,.\label{mlett:g405}\,\,\,
\eea
\eml 

Within the $\Delta$-bracket algebra the second-class constraints hold as 
strong identities. We now use this fact to eliminate from the game 
the variables $w^a, a=1,2$. As seen from (\ref{402}) and (\ref{403}), the
first-class constraints in the reduced phase space can be cast as 

\be
\label{406}
{\bar T}^{\ast}_a\,=\,p_a\,
+\,\frac{1}{2}\,m\,\omega\,\e_{ab}\,(\,q^b\,+\,y^b\,)\,\approx\,0\,\,\,.
\ee

\noindent
We subject, afterwards, the remaining variables to the transformation 

\bml
\label{407}
\bea
q^a\,\rightarrow\,{\bar q}^a\,& = &\,q^a\,\,\,,\label{mlett:a407}\\
p_a\,\rightarrow\,{\bar p}_a\,& = &\,p_a\,-\,\frac{1}{2}\,m\,\omega\,\e_{ab}\,y^b
\,\,\,,\label{mlett:b407}\\
y^a\,\rightarrow\,{\bar z} ^a\,& = &\,\sqrt{m \omega}\,y^a
\,\,\,.\label{mlett:c407}
\eea
\eml

\noindent
The variables ${\bar q}^a$, ${\bar p}_a$ are canonical, 

\be
\label{408}
\left[\,{\bar q}^a\,,\,{\bar p}_b\,\right]_{\Delta}\,= \,\delta^a_b \,\,\,,
\ee

\noindent
while

\be
\label{409}
\left[\,{\bar z}^a\,,\,{\bar z}^b\,\right]_{\Delta}\,= \,- \,\e^{ab}\,\,\,.
\ee

\noindent
All other $\Delta$-brackets vanish. In terms of these new variables, the
first-class constraints and the Hamiltonian of the reduced phase-space are
found to read, respectively, 

\be
\label{410}
{\bar T}^{\ast}_a\,=\,{\bar p}_a\,
+\,\e_{ab}\,\left (\frac{1}{2}\,m\,\omega\,\,{\bar q}^b\,
+\,\sqrt{m \omega}\,{\bar z}^b\,\right)\,\approx\,0\,\,\,,
\ee

\be
\label{411}
H^{\ast}_S\,= \,\frac{m \omega^2}{2}\,\left (\,{\bar q}^a\,
+\,\frac{1}{\sqrt{m \omega}} {\bar z}^a \right)^2\,\,\,.
\ee

From (\ref{401}), (\ref{402}), (\ref{410}) and (\ref{411}) one concludes that 
the BFT and the Stuckelberg embeddings lead to
equivalent results. The only subtle point to be noticed is that the
$z$'s are composite variables, while the ${\bar z}$'s are basic phase-space
variables. Nevertheless, the second-class constraints involving the 
${\bar z}$'s give rise to a Dirac bracket which is numerically equal to the 
Poisson bracket obeyed by the $z$'s. 

\section{BFT embedding of the self-dual model}
\label{sec:level5}

On a semiclassical level, the SD model has been shown\cite{Jackiw1,Frad} to 
be equivalent to the Maxwell-Chern-Simons (MCS) theory\cite{Ha1,Jackiw2,Gi2}. 
That this equivalence holds on the level of the Green functions was proved in
Ref.\cite{Rothe1}. Lately, the second-class constraints of the SD model were
successfully converted into first-class by means of the BFT embedding
procedure. It was then found that the SD model and the MCS theory in a Coulomb
like gauge are just different gauge-fixed versions of a parent 
theory\cite{Kim,Rothe2,Rothe3}. 

In this Section we go further on and construct explicitly, for any gauge, the 
phase space variables of the MCS theory in terms of those of the SD model. As 
we shall see, the strategy developed in Sections 2 and 3 will be of great help
for putting the equivalence between the SD and the MCS theories on a more 
firm basis.     

The dynamics of the SD theory is described by the Lagrangian
density\cite{TPN1,Jackiw1} 

\be
\label{501}
{\cal L}^{SD}\,=\,- \frac{1}{2\theta} \epsilon^{\mu\nu\rho}\,
(\partial_{\mu} f_{\nu})\,
f_{\rho}\,+\,\frac{1}{2}\,f^{\mu}f_{\mu}\,\,\,,
\ee

\noindent
where $\theta$ is a parameter with dimensions of mass. We use natural units 
($c=\hbar=1$) and our metric is $g_{00}=-g_{11}=-g_{22}=1$. The fully 
antisymmetric tensor $\e^{\mu\nu\rho}$ is normalized such that 
$\epsilon^{012}=1$ and we define $\epsilon^{ij}\equiv\epsilon^{0ij}$. Repeated
Greek indices sum from $0$ to $2$ while repeated Latin indices sum from $1$ to
$2$. Within the Hamiltonian framework, the SD model is characterized by the 
primary constraints\cite{Kim,Rothe2,Rothe3,Rothe1}  

\bml
\label{502}
\bea
&&{\cal T}_0^{(0)}\,=\,\pi_0\,\approx\,0\,\,\,, \label{mlett:a502} \\
&&{\cal T}_i^{(0)}\,=\,\pi_i\,
+\,\frac{1}{2\theta} \epsilon_{ij} f^j\,\approx\,0\,\,,i=1,2\,\,\,,
\label{mlett:b502}
\eea
\eml

\noindent
the secondary constraint

\be
\label{503}
{\cal T}_3^{(0)}\,=\,\frac{1}{\theta}\left(f^0\,
-\,\frac{1}{\theta}\epsilon_{ij}\partial^if^j\right)
\,\approx\,0\,\,\,,
\ee

\noindent
and the canonical Hamiltonian 

\be
\label{504}
H_{SD}^{(0)}\,=\,\int d^2x\,\left(-\frac{1}{2}f^{\mu}f_{\mu}\,
+\,\frac{1}{\theta} \e_{ij} f^0 \partial^i f^j \right)\,\,\,.
\ee

\noindent
We denote by $\pi_{\mu}$ the momentum canonically conjugate to the field
variable $f^{\mu}$. All constraints are second-class.

The quantization of the SD model as a second-class theory was carried out in 
detail in Ref.\cite{Gi3}. The Heisenberg equations of motion together with the
equal-time commutation relations are solved by

\be
\label{505}
f^{\mu(\pm)}(x)\,=\,\frac{1}{2\pi} \int \frac{d^2k}{\sqrt{2\omega_{\vk}}} 
\exp \left[\pm i(\omega_{\vk} x^0 - \vk \cdot
\vx)\right]\,f^{\mu(\pm)}(\vk)\,\,\,, 
\ee

\noindent
where $\omega_{\vk} \equiv +\sqrt{|\vk|^2+\theta^2}$ and

\bml
\label{506}
\bea
&&f^{\mu (+)}(\vk)\,=\,\ve^{\mu}(\vk)\,a^{(+)}(\vk)\,\,\,,\label{mlett:a506}\\
&&f^{\mu (-)}(\vk)\,=\,\ve^{\ast\mu}(\vk)\,a^{(-)}(\vk)\,\,\,.
\label{mlett:b506}
\eea
\eml 

\noindent
Here, $a^{(\pm)}(\vk)$ are creation and anhilation operators and 
$\ve^{\mu}(\vk)$ is the polarization vector. Observe that the system under
analysis possesses three coordinates, three momenta and four second-class
constraints. Thus, as in the particle case in Section 2, one is left with
only one independent degree of freedom. The determination of $\ve^{\mu}(\vk)$
led to\cite{Gi3}

\bml
\label{507}
\bea
\ve^{0}(\vk)\,& = &\, \frac{1}{|\theta|}\, \vk \cdot \vec{\ve}(0), 
\label{mlett:a507} \\
\ve^{j} (\vk)\,& = &\,\ve^{j}(0)\,+\,
\frac{\vec{\ve}(0) \cdot \vk}{(\omega_{\vk}\,+\,|\theta|)\,|\theta| }
\,k^{j}\,\,\,, \label{mlett:b507}
\eea
\eml  

\noindent
where

\bml
\label{508}
\bea
\ve^{0}(0)\,& = &\,0\,\,\,,\label{mlett:a508}\\
\ve^{j}(0)\,& = &\,-i\,\frac{|\theta|}{\theta}\,\e^{jl}\,\ve_{l}(0)\,\,\,.
\label{mlett:b508}
\eea
\eml
  
\noindent
As for the spin of the of the SD quanta, it was found to be $\pm 1$ depending
upon the sign of $\theta$. 

The outcomes of applying the BFT embedding procedure to the SD theory have 
already been reported in the literature and we shall merely quote here the
results\cite{Kim,Rothe2,Rothe3,Rothe1} 

\bml
\label{509}
\bea
&&{\cal T}_0^{(0)} \longrightarrow {\cal T}_0 \,=\,\pi_0\,-\,\frac{1}{\theta}\,
\phi^0 \,\approx\,0\,\,\,,\label{mlett:a509}\\
&&{\cal T}_i^{(0)} \longrightarrow {\cal T}_i \,=\,\pi_i\,+\,\frac{1}{\theta}\,
\e_{ij}\,\left( \frac{1}{2}\,f^j + \phi^j \right)\,-\,\frac{1}{\theta^2}\,\e_{ij}
\,\pa_j \phi^0\,\approx\,0\,\,\,,\label{mlett:b509}\\
&&{\cal T}_3^{(0)} \longrightarrow {\cal T}_3 \,=\,\frac{1}{\theta}\,
\left( f^0 + \phi^3 \right)\,-\,\frac{1}{\theta^2}\,\e_{ij}\,\pa^i \left(f^j + 
\phi^j \right) \,\approx\,0\,\,\,,\label{mlett:c509}
\eea
\eml

\noindent
and

\be
\label{510}
H_{SD}^{(0)} \longrightarrow H_{SD}\,
=\,\int d^2x\,\left[\,\frac{1}{2}\,\left( f^i + 
\phi^i \right)^2\,+\,\frac{1}{2}\,\left( f^0 +  \phi^3 \right)^2\,-\,
m\,\left( f^0 +  \phi^3 \right)\,{\cal T}_3 \right]\,\,\,.
\ee

\noindent
As demanded\cite{BFT}, a new pair of canonical phase-space variables 
(coordinates $u^0$, $u^i$, $u^3$ and momenta ${\cal P}_0$, ${\cal P}_i$, 
${\cal P}_3$) for each second-class constraint has been introduced. By 
definition

\bml
\label{511}
\bea
&&\phi^0\,\equiv \,-\,\frac{1}{2}\,u^0\,-\,\theta\,{\cal P}_3
\,\,\,,\label{mlett:a511}\\
&&\phi^i\,\equiv \,-\,\frac{1}{2}\,u^i\,+\,\pa^i\,{\cal P}_3\,
-\,\theta\,\e^{ij}\,{\cal P}_j\,\,\,,\label{mlett:b511}\\
&&\phi^3\,\equiv \,-\,\frac{1}{2}\,u^3\,+\,\theta\,{\cal P}_0\,
+\,\pa^i{\cal P}_i\,\,\,.\label{mlett:c511}
\eea
\eml 
  
\noindent
Then, the only nonvanishing Poisson brackets among the $\phi$'s are, as 
required\cite{BFT},

\bml
\label{512}
\bea
&&\left[\phi^0(\vx)\,,\,\phi^3(\vy)\,\right]_P\,=\,-\,\theta\,\dxy\,\,\,,
\label{mlett:a512}\\
&&\left[\phi^i(\vx)\,,\,\phi^j(\vy)\,\right]_P\,=\,
-\,\theta\,\e^{ij}\,\dxy\,\,\,,\label{mlett:b512}\\
&&\left[\phi^3(\vx)\,,\,\phi^i(\vy)\,\right]_P\,=\,-\,\pa_i^x\,\dxy\,\,\,.
\label{mlett:c512}
\eea
\eml

Since the extended constraints verify, by construction, an Abelian 
algebra\cite{BFT}, the Green functions generating functional 
(${\cal W}_{\chi}$) is given by 

\bea
\label{513}
{\cal W}_{\chi}\,&=&\,{\cal N}\,\int\,[\prod_{\mu=0}^2 {\cal D}f^{\mu}]\,
[\prod_{\mu=0}^2 {\cal D}\pi_{\mu}]\,
[\prod_{a=0}^3 {\cal D}u^a]\,[\prod_{a=0}^3 {\cal D}{\cal P}_a]\,
\det[\chi^a,{\cal T}_b]_P\nonumber\\
&\times &
\left(\prod_{a=0}^{3}\,\delta [\,{\cal T}_a\,]\right)
\left(\prod_{a=0}^{3}\,\delta [\,\chi^a\,]\right)
\exp\left[\,i\,\int \,d^3x\,
\left( \pi_{\mu} {\dot f}^{\mu}\,+\, {\cal P}_{a} {\dot u}^{a}\, 
-\,{\cal H}_{SD}\,\right)\right]\,\,\,,
\eea

\noindent
where ${\cal H}_{SD}$ is the Hamiltonian density corresponding to $H_{SD}$ and
$\chi^a,
a=0,1,2,3$ are the gauge conditions. As we did in the particle case, we first
perform the change variables $u^a \rightarrow u^{\prime a} = \phi^a$, 
${\cal P}_a \rightarrow {\cal P}^{\prime}_{a} = {\cal P}_a$, whose jacobian is
nonsingular and can be lumped into the normalization constant ${\cal N}$. 
Since ${\cal H}_{SD}$ does not depend upon ${\cal P}^{\prime}_{a}$, the 
corresponding integrals can be carried out at once, yielding,

\bea
\label{514}
{\cal W}_{\chi}\,&=&\,{\cal N}\,\int\,[\prod_{\mu=0}^2 {\cal D}f^{\mu}]\,
[\prod_{\mu=0}^2 {\cal D}\pi_{\mu}]\,
[\prod_{a=0}^3 {\cal D}\phi^a]\,
\det[\chi^a,{\cal T}_b]_P\,
\left(\prod_{a=0}^{3}\,\delta [\,{\cal T}_a\,]\right)
\left(\prod_{a=0}^{3}\,\delta [\,\chi^a\,]\right)
\nonumber\\
&\times & \exp\left[\,i\,\int \,d^3x\,
\left( \pi_{\mu} {\dot f}^{\mu}\,
+\,\frac{1}{2\theta}\,\phi^i \e_{ij} {\dot \phi}^j
\,-\,\frac{1}{\theta}\,\phi^3 {\dot \phi}^0
\,-\,\frac{1}{\theta^2}\,\phi^0 \e_{ij} \pa^i {\dot \phi}^j\, 
-\,{\cal H}_{SD}\,\right)\right]\,\,\,.
\eea

\noindent
We use next the extended primary constraints ${\cal T}_0 = 0, 
{\cal T}_i = 0$ to perform the momentum integrals, thus arriving to

\bea
\label{515}
{\cal W}_{\chi}\,&=&\,{\cal N}\,\int\,[\prod_{\mu=0}^2 {\cal D}f^{\mu}]\,
[\prod_{a=0}^3 {\cal D}\phi^a]\,
\det[\chi^a,{\cal T}_b]_P\,
\left(\prod_{a=0}^{3}\,\delta [\,\chi^a\,]\right)
\nonumber\\
& \times & \delta \left[ \frac{1}{\theta}\left(C^0\,
-\,\frac{1}{\theta}\epsilon_{ij}\partial^iC^j\right) \right]
\exp\left[\,i\,\int \,d^3x\, {\cal L}^{SD}(C^{\mu}) \right]\,\,\,.
\eea

This is the analog of (\ref{314}). The role of the nonrelativistic variables 
$x^a$ (see Eq.(\ref{316})) is now played by the variables $C^{\mu}$, defined 
as 

\bml
\label{516}
\bea
&& C^0\,=\,f^0\,+\,\phi^3\,\,\,,\label{mlett:a516}\\
&& C^i\,=\,f^i\,+\,\phi^i\,\,\,.\label{mlett:b516}
\eea
\eml

\noindent
The variables $C^{\mu}$ are gauge invariant. To see this, we recall
that the generator of infinitesimal gauge transformations ($G$) is, by
definition\cite{Di1}, a linear combination of the first-class constraints,
i.e.,

\be
\label{517}
G\,=\,\int\,d^2x\,\ve^a\,{\cal T}_a\,\,\,,
\ee

\noindent
where the $\ve^a$ are infinitesimal gauge parameters and $a$ runs from $0$ 
to $3$. Then, under infinitesimal gauge transformation the phase space
variables can be seen to change as follows

\bml
\label{518}
\bea
& & \delta f^0\,=\,\ve^0\,\,\,,\label{mlett:a518}\\
& & \delta f^i\,=\,\ve^i\,\,\,,\label{mlett:b518}\\
& & \delta u^0\,=\,\ve^3\,\,\,,\label{mlett:c518}\\
& & \delta u^i\,=\,\ve^i\,\,\,,\label{mlett:d518}\\
& & \delta u^3\,=\,\ve^0\,\,\,,\label{mlett:e518}\\
& & \delta \pi_0\,=\,-\,\frac{1}{m}\,\ve^3\,\,\,,\label{mlett:f518}\\
& & \delta \pi_i\,=\,\frac{1}{2 m}\,\e_{ij}\,\ve^j\,+\,\frac{1}{m^2}\,
\e_{ij}\,\pa^j \ve^3 \,\,\,,\label{mlett:g518}\\
& & \delta {\cal P}_0\,=\,-\,\frac{1}{2 m}\,\ve^0\,+\,\frac{1}{2 m^2}\,
\e_{ij}\,\pa^i\ve^j \,\,\,,\label{mlett:h518}\\
& & \delta {\cal P}_i\,=\,-\,\frac{1}{2 m}\,\e_{ij}\,\ve^j\,
-\,\frac{1}{2 m^2}\,\e_{ij}\,\pa^j\ve^3 \,\,\,,\label{mlett:i518}\\
& & \delta {\cal P}_3\,=\,\frac{1}{2 m}\,\ve^3\,\,\,.\label{mlett:j518}
\eea
\eml

\noindent
From (\ref{516}) and (\ref{518}) follows that the $C^{\mu}$'s are, as asserted,
gauge invariant fields. Moreover, the subsidiary conditions 
$\chi^a = \phi^a = 0$ return us back to the original second-class theory and,
hence, define the unitary gauge.

We turn next into investigating the existence of alternative formulations for
the converted theory. Guided by the particle case, we perform the canonical 
transformation

\bml
\label{519}
\bea
&& f^0\,\rightarrow\,A^0\,=\,f^0\,\,\,,\label{mlett:a519}\\
&& \pi_0\,\rightarrow\,P_0\,=\,\pi_0\,+\,\frac{1}{2\theta}\,u^0\,+\,{\cal P}_3
\,\,\,,\label{mlett:b519}\\
&& f^i\,\rightarrow\,A^i\,=\,\frac{1}{2}\,\left(f^i - u^i \right)\,+\,
\theta\,\e^{ij}\left(\pi_j + {\cal P}_j \right)\,
+\,\frac{1}{\theta}\,\pa^i u^0\,\,\,,
\label{mlett:c519}\\
&& \pi_i\,\rightarrow\,P_i\,=\,\frac{1}{2}\,\left(\pi_i - {\cal P}_i \right)\,
-\,\frac{1}{4\theta}\,\e_{ij}\left(f^j + u^j \right)\,-\,\frac{1}{\theta}\,
\e_{ij} \pa^j {\cal P}_3\,\,\,,\label{mlett:d41sy5}\\
&& u^0\,\rightarrow\,U^0\,=\,u^0\,\,\,,\label{mlett:e519}\\
&& {\cal P}_0\,\rightarrow\,N_0\,=\,{\cal P}_0\,+\,\frac{1}{2\theta}\,f^0\,+\,
\frac{1}{4\theta}\,\pa^i \left(\pi_i - {\cal P}_i\right)\,
-\,\frac{3}{8 \theta^2}
\e_{ij}\,\pa^i \left(f^j + u^j\right)\,\,\,,\label{mlett:f519}\\
&& u^i\,\rightarrow\,U^i\,=\,\frac{1}{2}\,\left(u^i + f^i \right)\,-\,
\theta\,\e^{ij}\left(\pi_j - {\cal P}_j \right)\,-\,3\,\pa^i {\cal P}_3\,\,\,,
\label{mlett:g519}\\
&& {\cal P}_i\,\rightarrow\,N_i\,=\,\frac{1}{2}\,{\cal T}_i 
\,\,\,,\label{mlett:h519}\\
&& u^3\,\rightarrow\,U^3\,=\,u^3\,-\,f^0\,+\,
\frac{1}{2}\,\pa^i \left(\pi_i + {\cal P}_i\right)\,+\,\frac{5}{4\theta}
\e_{ij}\,\pa^i \left(f^j - u^j\right)\,\,\,,\label{mlett:i519}\\
&& {\cal P}_3\,\rightarrow\,N_3\,=\,{\cal P}_3\,\,\,.\label{mlett:j519}
\eea
\eml

\noindent
In terms of the new variables, the functional integral in the right hand side
of (\ref{513}) can be cast as

\bea
\label{520}
{\cal W}_{\chi}\,&=&\,{\cal N}\,\int\,[\prod_{\mu=0}^2 {\cal D}A^{\mu}]\,
[\prod_{\mu=0}^2 {\cal D}P_{\mu}]\,
[\prod_{a=0}^3 {\cal D}U^a]\,[\prod_{a=0}^3 {\cal D} N_a]\,
\det[\chi^a,{\cal T}_b]_P\nonumber\\
&\times &
\left(\prod_{a=0}^{3}\,\delta [\,{\cal T}_a\,]\right)
\left(\prod_{a=0}^{3}\,\delta [\,\chi^a\,]\right)
 \exp\left[\,i\,\int \,d^3x\,
\left( P_{\mu} {\dot A}^{\mu}\,+\, N_{a} {\dot U}^{a}\, 
-\,{\cal K}\,\right)\right]\,\,\,,
\eea

\noindent
where the transformed Hamiltonian density ${\cal K}$ is given by

\bea
\label{521}
{\cal K}\,&=&\,\frac{\theta^2}{2}\,P_i P_i\,-\,\frac{\theta}{2}\,P_i \e^{ij} A^j\,+\,
\frac{1}{4 \theta^2}\,F_{ij} F^{ij}\,+\,\frac{1}{8}\,A^i A^i\nonumber\\
&-& \frac{1}{2 \theta}\,\e_{ij}\,F^{ij}\,{\cal G}\,-\,{\cal G}^2\,\,\,,
\eea

\noindent
with

\be
\label{522}
F^{ij}\,=\,\pa^i A^j - \pa^j A^i\,\,\,,
\ee

\noindent
and

\be
\label{523}
{\cal G}\,=\,\pa^k P_k\,+\,\frac{1}{2\theta}\,\e_{kl}\,\pa^k A^l\,\,\,.
\ee

\noindent
Also notice that, in terms of the new variables, the constraints translate
into  

\bml
\label{524}
\bea
&& {\cal T}_0\,=\,P_0\,\,\,,\label{mlett:a524}\\
&& {\cal T}_i\,=\,2\,N_i\,\,\,,\label{mlett:b524}\\
&& {\cal T}_3\,=\,N_0\,-\,\frac{1}{2\theta}\,U^3\,
-\,\frac{1}{2\theta}\,{\cal G} \,\,\,.\label{mlett:c524}
\eea
\eml

We now focus on the right hand side (\ref{520}). Since ${\cal K}$ does not 
depend upon $U^a$ and $N_a$, the partial gauge fixing $\chi^i = U^i = 0$ does 
not imply in a physically meaningful restriction and enables us to carry out 
the $U^i$ and $N_i$ integrals. The situation is quite analogous to that 
encountered in the particle case. Afterwards, the constraint ${\cal T}_3$ is
exponentiated by means of the auxiliary variable $\lambda^3$ and the integrals
on $N_0$ and $U^3$ are performed. The integrals on $N_3$ and $\lambda^3$ are
also carried out and one arrives at  

\bea
\label{525}
{\cal W}_{\chi}\,&=&\,{\cal N}\,\int\,[\prod_{\mu=0}^2 {\cal D}A^{\mu}]\,
[\prod_{\mu=0}^2 {\cal D}P_{\mu}]\,
[{\cal D}U^0]\,
\det[\chi^{a^{\prime}},{\cal T}_{b^{\prime}}]_P\nonumber\\
&\times &
\delta [P_0]\,\delta [\,\chi^0\,]\,\delta [\,\chi^3\,]
 \exp\left[\,i\,\int \,d^3x\,
\left( P_{\mu} {\dot A}^{\mu}\,-\,U^0 \,{\dot {\cal G}}\, 
-\,{\cal K}\,\right)\right]\,\,\,,
\eea

\noindent
where $a^{\prime}$ and $b^{\prime}$ only take the values $0$ and $3$. The 
subsequent integration on $U^0$ produces $\delta[{\dot {\cal G}}]$ which, up 
to a field independent determinant, is proportional to $\delta[{\cal G}]$. 
Hence, the final form for ${\cal W}_{\chi}$ is

\bea
\label{526}
&&{\cal W}_{\chi}\,=\,{\cal N}\,\int\,[\prod_{\mu=0}^2 {\cal D}A^{\mu}]\,
[\prod_{\mu=0}^2 {\cal D}P_{\mu}]\,
\det[\chi^{a^{\prime}},{\cal T}_{b^{\prime}}]_P
\delta [P_0]\,\delta [\,\chi^0\,]\,
\delta[ \pa^k P_k + \frac{1}{2 \theta} \e_{kl} \pa^k A^l]\,\delta [\,\chi^3\,]
\nonumber\\ 
& \times & \exp\left\{\,i\,\int \,d^3x\,
\left[ P_{\mu} {\dot A}^{\mu}\, 
-\left(\,\frac{\theta^2}{2}\,P_i P_i\,-\,\frac{\theta}{2}\,P_i \e^{ij} A^j\,+\,
\frac{1}{4 \theta^2}\,F_{ij} F^{ij}\,+\,\frac{1}{8}\,A^i A^i\,\right)\right]
\right\}\,\,\,,
\eea 

\noindent
which is just the phase-space path integral describing the MCS theory in any
arbitrary canonical gauge\cite{Gi2}. Also, from Eqs.(\ref{518}) and 
(\ref{519}) one can confirm that, under infinitesimal gauge 
transformations, the field variables $A^{\mu}$ and $P_{\mu}$ do transform as 
the MCS variables, i.e., 

\bml
\label{527}
\bea
&& \delta A^0 \,=\,\ve^0\,\,\,,\label{mlett:a527}\\
&& \delta P_0 \,=\,0\,\,\,,\label{mlett:b527}\\
&& \delta A^i \,=\,\frac{1}{2\theta}\,\pa^i \ve^3\,\,\,,\label{mlett:c527}\\
&& \delta P_i \,=\,\frac{1}{4\theta^2}\,\e_{ik} \pa^k \ve^3
\,\,\,.\label{mlett:d527}
\eea
\eml 

After performing the momentum integrations, one finds that the effective 
Lagrangian arising 
from (\ref{527}) is the well known MCS Lagrangian density 
${\cal L}^{MCS}$\cite{Ha1,Jackiw2,Gi2}

\be
\label{528}
{\cal L}^{MCS}\,\equiv\,{\cal L}_{U^i=0}\,
=\,-\frac{1}{4 \theta^2}F_{\mu\nu}F^{\mu\nu}
+\frac{1}{4\theta} \epsilon^{\mu\nu\alpha}F_{\mu\nu}A_{\alpha}\,\,\,.
\ee

\noindent
As in the particle case, the mass term in (\ref{501}) was replaced by a 
``kinetic energy'' term. However, unlike in the particle case, 
${\cal L}^{MCS}$ can not be fully written in terms of gauge
invariant fields. In fact, ${\cal L}^{MCS}$ does not describe a regular
theory but a first-class one, possessing two first-class constraints\cite{Gi2}.
The counting of degrees of freedom (three coordinates, three momenta, two
first-class constraints and two gauge conditions) reveals that, as in the SD
model, only one independent degree of freedom is present in the MCS theory. 
Since ${\cal L}^{MCS}$ derives from ${\cal L}^{SD}$ through the BFT conversion
procedure, they must be equivalent. In fact, as demonstrated in
Ref.\cite{Gi3}, the particle content of the SD model is that of the MCS
model in the Coulomb gauge, while the polarization vector of the SD quanta (see
Eqs.(\ref{507}) and (\ref{508})) is that of the massive MCS quanta in the 
Landau gauge\footnote{Recall that in the Landau gauge the MCS theory exhibits 
massive and massless (gauge) excitations.}. It is in this sense that 
${\cal L}^{MCS}$ and ${\cal L}^{SD}$ describe equivalent physics.   

We believe to have generalized the proof of equivalence between the SD and the
MCS models presented in Refs.\cite{Kim,Rothe2,Rothe3,Rothe1}, where the 
election of specific functional forms for $\chi^0$ and $\chi^3$ was essential
to establish this correspondence. Notice, moreover, that our proof of
equivalence is not restricted to demonstrate the equality between two
functional integrals, but also involves an explicit construction of the 
phase-space variables of the MCS theory in terms of those of the SD theory. 
This is precisely the meaning of the canonical transformation in 
Eq.(\ref{519}).  

\section{BFT embedding of the Proca-Wentzel theory coupled to fermions}
\label{sec:level6}

We shall present in this Section the BFT embedding of the $3+1$-dimensional 
Proca-Wentzel field ($B^{\alpha}$) minimally coupled to fermions 
($ \p $, $ \bp $)\footnote{The BFT embedding of the free Proca-Wentzel field 
has already been reported in the literature. See 
Refs.\cite{Yoon1,Yoon2,Ban2}.}. This is 
second-class theory involving bosonic and fermionic degrees of freedom and 
exhibiting bosonic and fermionic constraints. Our starting point is
the Lagrangian density 

\be
\label{601}
{\cal L}^{PWF} = -\frac{1}{4}F^B_{\alpha\beta}F^{B,\alpha\beta}
+ \frac{\mu^2}{2}\,B^{\alpha}B_{\alpha}
+\frac{i}{2}\bar{\psi}\gamma^{\alpha}\partial_{\alpha}\psi-
\frac{i}{2}(\partial_{\alpha}\bar{\psi})\gamma^{\alpha}\psi-M\,\bar{\psi}\psi
+g\,\bar{\psi}\gamma^{\alpha}B_{\alpha}\psi ,
\ee

\noindent
where $F^B_{\alpha\beta} = \pa_{\alpha} B_{\beta}-\pa_{\beta} B_{\alpha}$,
$\mu$ is the vector boson mass, $M$ is the fermion mass and $g$ is a coupling
constant. Our metric is $g_{00}=-g_{11}=-g_{22}=-g_{33}=+1$. This model
possesses primary and secondary second-class bosonic constraints. They are,
respectively, 

\bml
\label{602}
\bea
b_0^{(0)}\,&=&\,\pi^B_0\,\approx \,0\,\,\,,\label{mlett:a602}\\
b_1^{(0)}\,&=&\,\frac{1}{m}\,(\,\pa^i \pi^B_i\,+\,m^2\,B^0\,
+\,g\,\bp \g^0 \p\,)\,\approx \,0\,\,\,.\label{mlett:b602}
\eea
\eml

\noindent
The second-class fermionic constraints,
 
\bml
\label{603}
\bea
f_a^{(0)}\,&=&\,\pi_{\bp_a}\,-\,\frac{i}{2}\,\g^0_{ab}\,\p_b\,\approx \, 0\,
\,\,,\label{mlett:a603}\\
{\bar f}_a^{(0)}\,&=&\,\pi_{\p_a}\,
-\,\frac{i}{2}\,\bp_b\,\g^0_{ba}\,\approx \, 0\,\,\,,\label{mlett:b603}
\eea
\eml

\noindent
are all primary constraints, while the canonical Hamiltonian reads  

\bea
\label{604}
H_{PWF}^{(0)}\, =  \int d^3x \,& & \left\{ \frac {1}{2} \pi^B_i\pi^B_i\,
+\, \frac {1}{4}F^B_{ij} F^{B,ij}\,+\,\frac{\mu^2}{2}\,B^i B^i\,
 + \frac{i}{2} \left[\pa_i \bp\right] \g^i \p \,-
\,\frac{i}{2}\,\bp\, \g^i \left[\pa_i\p\right] \right. \nonumber \\ 
&& \left.
\,+\,g \,\bp \g^i \p\, B^i\,
+\,M \bp\p \,-\,B^0\,(\,\pa^i \pi^B_i\,+\,\frac{\mu^2}{2}\,B^0\,
+\,g\,\bp \g^0 \p\,)\right \}\,\,\,.
\eea

\noindent
Here, $\pi^B_{\mu}$, $\pi_{\p_a}$ and $\pi_{\bp_a}$ are the momenta canonically
conjugate to $B^{\mu}$, $\p_a$ and $\bp_a$, respectively, and $a$ is a spinor
index running from $1$ to $4$. Notice that 
${\bar f}^{(0)} = -f^{\dagger}\g^0$. 

Now, we use again the BFT procedure to convert the system into first-class. 
For the bosonic sector of the constraints one obtains 

\bml
\label{605}
\bea
b_0^{(0)}\,\rightarrow \,b_0 \,&=&\,b_0^{(0)}\,+\,b_0^{(1)}\,=\,
\pi_0\,+\,m\,\Phi^1\,\approx \,0\,\,\,,\label{mlett:a605}\\
b_1^{(0)}\,\rightarrow \,b_1 \,&=&\,b_1^{(0)}\,+\,b_1^{(1)}+\,b_1^{(2)}
\nonumber\\
&=& \frac{1}{m}\,\left[\,\pa^i \pi^B_i\,+\,\mu^2\,(B^0 - \Phi^0 )\,
+\,g\,(\bp + \frac{1}{2}\bx ) \g^0 
\,(\p + \frac{1}{2} \x) \, \right]
\,\approx \,0\,\,\,,\label{mlett:b605}
\eea
\eml

\noindent
whereas the fermionic sector of the constraints extends as follows

\bml
\label{606}
\bea
f_a^{(0)}\,\rightarrow \,f_a \,&=&\,f_a^{(0)}\,+\,f_a^{(1)}\,=\,
\pi_{\bp_a}\,-\,\frac{i}{2}\,\g^0_{ab}\,(\p_b+\x_b)\,\approx \, 0\,
\,\,,\label{mlett:a606}\\
{\bar f}_a^{(0)}\,\rightarrow \,{\bar f}_a \,&=&\,{\bar f}_a^{(0)}\,
+\,{\bar f}_a^{(1)}\,=\,
\pi_{\p_a}\,-\,\frac{i}{2}\,(\bp_b+\bx_b)\,\g^0_{ba}\,\approx \, 0\,
\,\,.\label{mlett:b606}
\eea
\eml

\noindent
The extension of the Hamiltonian is more involved and one ends up with

\bea
\label{607}
&&H_{PWF}^{(0)} \rightarrow H_{PWF}\,=\,H_{PWF}^{(0)}\,+\,H_{PWF}^{(1)}\,
+\,H_{PWF}^{(2)}\,+\,H_{PWF}^{(3)}
\nonumber\\
& = & \int d^3x \,\left\{ \frac {1}{2} \pi^B_i\pi^B_i\,
+\, \frac {1}{4}F^B_{ij} F^{B,ij}\,+\,\frac{\mu^2}{2}\,B^i B^i\,
+\,\frac{\mu^2}{2}\,(B^0 - \Phi^0)^2\,-\,\mu\,(B^0 - \Phi^0)\,b_1
\right. \nonumber\\ 
&  + & \left. 
\frac{i}{2}\,\left[\pa_i (\bp + \frac{1}{2} \bx)\right] 
\g^i (\p + \frac{1}{2} \x) \,- \,\frac{i}{2}\,(\bp + \frac{1}{2} \bx)\, 
\g^i \left[ \pa_i (\p + \frac{1}{2} \x)\right] 
\right.\nonumber\\
& + & \left.
\,g \,(\bp + \frac{1}{2} \bx) \g^i (\p + \frac{1}{2} \x)\, B^i\,
\,+ \,M (\bp + \frac{1}{2} \bx)(\p + \frac{1}{2} \x) \, 
+\,m\,\Phi^1\,\pa_iB^i\,-\,\frac{1}{2}\,\Phi^1 \nabla^2 \Phi^1 \right\}\,\,\,,
\eea

\noindent
where we denoted by $\Phi^0$, $\Phi^1$ and $\bx$, $\x$ the BFT bosonic and 
fermionic embedding variables, respectively. Presently, we omit the detailed 
construction of these composite objects in terms of canonical variables but
only mention that they are required to obey the Poisson bracket relations

\bml
\label{608}
\bea
\left[\,\Phi^A(x^0,\vx)\,,\,\Phi^B(x^0,\vy)\,\right]_P\,& = &\,\frac{1}{\mu}\,
\e^{AB}\,\dxy\,\,\,, \label{mlett:a608}\\
\left[\,\x_a(x^0,\vx)\,,\,\bx_b(x^0,\vy)\,\right]_P\,& = &\,-4i\,
\g^0_{ab}\,\dxy\,\,\,, \label{mlett:b608}
\eea
\eml

\noindent
where the superscripts $A$ and $B$ run from $0$ to $1$. Needless to say, the
Poisson bracket between fermionic variables is symmetric. We would like to 
stress that, as a consequence of the embedding procedure, the number of 
fermions doubles\footnote{We are indebted to Prof. J Barcelos-Neto for a 
discussion about this effect.}.
One can check that the constraints $b_A$, $f_a$, ${\bar f}_a$ and $H$ are 
strong under involution, thus characterizing an Abelian first-class theory. 
By definition, the Hermitean generator of infinitesimal gauge transformations 
is

\be
\label{609}
G_{PWF}\,=\,\int\,d^3x\,\left( \lambda^A b_A\,+\,i\,{\bar \alpha}^a f_a\,+\,
i\,{\bar f}_a \alpha^a\right)\,\,\,,
\ee

\noindent
where $\lambda_A$, $\alpha^a$ and ${\bar \alpha}^a = + \alpha^{\dagger}
\g^0$ are space-time dependent
gauge parameters. Then, under infinitesimal gauge transformations the field 
in the game change as 

\bml
\label{610}
\bea
\delta B^{0} \,& = &\,\lambda^0\,\,\,,\label{mlett:a610}\\
\delta B^{i} \,& = &\,-\,\frac{1}{m}\,\pa^i \lambda^1\,\,\,,
\label{mlett:b610}\\
\delta \Phi^0 \,& = &\,\lambda^0\,\,\,,\label{mlett:c610}\\
\delta \Phi^1 \,& = &\,\lambda^1\,\,\,,\label{mlett:d610}\\
\delta \p_a \,& = &\,i\,\alpha^a\,\,\,,\label{mlett:e610}\\
\delta \bp_a \,& = &\,-\,i\,{\bar \alpha}^a\,\,\,,\label{mlett:f610}\\
\delta \x_a \,& = &\,-\, \frac{2 i g}{\mu}\,(\p_a 
+ \frac{1}{2} \x_a)\,\lambda^1
\,-\,2\,i\,\alpha^a\,\,\,,\label{mlett:g610}\\
\delta \bx_a \,& = &\, +\,\frac{2 i g}{\mu}\,(\bp_a 
+ \frac{1}{2} \bx_a)\,\lambda^1
\,+\,2\,i\,{\bar \alpha}^a\,\,\,.\label{mlett:h610}
\eea
\eml

We look next for the generating functional of Green functions 
$W_{\chi,\eta, {\bar \eta}}$. Since the theory is Abelian, the ghosts are 
easily integrated out and one is left with

\bea
\label{611}
W_{\chi,\eta, {\bar \eta}}\,
&=&\,{\cal N}\,\int\,[D\Sigma]\,\det [\chi^A, b_A]\,
\det [{\bar \eta}^a, f_a]\,\det [\eta^a, {\bar f}_a]\,
\left(\prod_{A=0}^{1}\,\delta [\,b_A\,]\right)
\left(\prod_{A=0}^{1}\,\delta [\,\chi^A\,]\right)\nonumber\\
&&\left(\prod_{a=1}^{4}\,\delta [\,f_a\,]\right)
\left(\prod_{a=1}^{4}\,\delta [\,{\bar \eta}^a\,]\right)
\left(\prod_{a=1}^{4}\,\delta [\,{\bar f}_a\,]\right)
\left(\prod_{a=1}^{4}\,\delta [\, \eta^a\,]\right) \,
\exp \left(i {\cal A}_{PWF}\right)\,\,\,,
\eea

\noindent
where $\chi^A$ are the gauge fixing functions for the bosonic sector of
constraints and $\eta^a$ and ${\bar \eta}^a$ are the corresponding ones for the
fermionic sector. These subsidiary conditions are to be chosen as to make all
the Faddeev-Popov determinants, in Eq.(\ref{611}), nonvanishing and 
field-independent. The integration measure $[D\Sigma]$ involves all variables
appearing in the action ${\cal A}_{PWF}$, which, in turn, reads

\be
\label{612}
{\cal A}_{PWF}\,=\,\int\,d^4x\,\left( \pi^B_{\mu}{\dot B}^{\mu}\,
+\,\pi_{\p}{\dot \p}
\,- {\dot {\bp}} \pi_{\bp}\,+\,\frac{i}{4} \bx \g^0 {\dot \x}\,+\,
\frac{\mu}{2}\,\Phi^A \e_{AB} {\dot \Phi}^B\,-\,{\cal H}_{PWF}\right)\,\,\,,
\ee

\noindent
with $\int d^3x {\cal H}_{PWF} = H_{PWF}$. The integrations over the fermionic
momenta can be carried out at once by using $f_a=0$ and ${\bar f}_a=0$. Then,
the gauge 
is partially fixed by choosing $\chi^0=B^0$. This, together with 
$b_0=0$, enables one to integrate out the sector $B^0$, $\pi^B_0$. Afterwards, 
the constraint $b_1=0$ is exponentiated by means of the Lagrange multiplier
$\mu B^0$, which brings back $B^0$ into the game. Finally, by 
restricting
the remaining gauge conditions not to depend upon $\Phi^0$ and the momenta
$\pi^B_i$, the corresponding integrals can also be carried out. Thus, one 
arrives at    
 
\bea
\label{613}
W_{\chi,\eta,{\bar \eta}}\,&=&\,{\cal N}\,\int\,[\prod_{\alpha} DB^{\alpha}]
[D\theta][D\bp][D\p][D\bx][D\x]\, \det [\chi^1, b_1]\,
\det [{\bar \eta}^a, f_a]\,\det [\eta^a, {\bar f}_a]\,\nonumber\\
&&\delta[\chi^1]\,
\left(\prod_{a=1}^{4}\,\delta [\,{\bar \eta}^a\,]\right)\,
\left(\prod_{a=1}^{4}\,\delta [\, \eta^a\,]\right)\,
\exp\left(i \int\,d^4x\,{\cal L}^{PWF}_{BFT}\right)\,\,\,,
\eea

\noindent
where

\bea
\label{614}
{\cal L}^{PWF}_{BFT}\,& =  & -\frac{1}{4}\,F^B_{\alpha\beta}F^{B,\alpha\beta}
+ \frac{m^2}{2}\,\left( B^{\alpha} - \pa^{\alpha} \theta \right)
\left( B_{\alpha} - \pa_{\alpha} \theta \right)\nonumber\\
&+&\frac{i}{2}(\bp + \frac{1}{2}\bx) \gamma^{\alpha}
\left[ D_{\alpha}(B) (\p + \frac{1}{2} \x)\right] -
\frac{i}{2}\left[D^{\ast}_{\alpha}(B) ( \bp + \frac{1}{2}\bx)\right]
\gamma^{\alpha}\,(\p + \frac{1}{2}\x)\nonumber\\ 
&-& M\,(\bp + \frac{1}{2}\bx) 
(\p + \frac{1}{2}\x)\,\,\,.
\eea

\noindent
As usual, $D_{\alpha}(B)\,\equiv\,\pa_{\alpha}\,-\,ig\,B_{\alpha}$ and we have
replaced $\Phi^1$ by $ - \mu \theta$.

From (\ref{610}) follows that ${\cal L}^{PWF}_{BFT}$ remains invariant under
gauge transformations. Clearly, the unitary 
gauge conditions $\chi^1 = \theta = 0$, ${\bar \eta} = 
\bx = 0$ and $\eta = \x =0$ lead us back to the original theory, defined by
${\cal L}^{PWF}$.  

We end this work by remarking that, as in the particle case, the theory 
can be fully phrased in terms of 
local gauge invariant variables. To see how this come about, we start by 
introducing the composite fields $\Psi$, ${\bar \Psi}$ and ${\cal B}^{\alpha}$
via the non-linear transformation  

\bml
\label{615}
\bea
\Psi\,& \equiv &\,e^{-ig\theta}\,\left(\p + \frac{1}{2} \x\right)
\,\,\,,\label{mlett:a615}\\
{\bar \Psi}\, &\equiv &\,e^{+ig\theta}\,\left(\bp + \frac{1}{2} \bx\right)
\,\,\,,\label{mlett:b615}\\
{\cal B}^{\alpha}\,& \equiv &\,B^{\alpha}\,-\,\pa^{\alpha}\theta
\,\,\,,\label{mlett:c615}
\eea
\eml

\noindent
which, from (\ref{610}), are seen to be effectively gauge invariant. It is now
easy to see that ${\cal L}^{PWF}_{BFT}$ can be casted as 

\bea
\label{616}
{\cal L}^{PWF}_{BFT}\,&=&\,
-\,\frac{1}{4}\,F^{\cal B}_{\alpha \beta}F^{{\cal B},\alpha
\beta}\, +\,\frac{\mu^2}{2}\,{\cal B}^{\alpha}{\cal B}_{\alpha}\nonumber\\
&+&\frac{i}{2}\,{\bar \Psi}\g^{\alpha} 
\left[D_{\alpha}({\cal B}) \Psi\right]\,
- \,\frac{i}{2}\,\left[D^{\ast}_{\alpha}({\cal B}){\bar
\Psi}\right]\g^{\alpha}\Psi\,-\,M\,{\bar \Psi}\Psi\,\,\,.
\eea  

Therefore, the theories described by ${\cal L}^{PWF}$ and ${\cal
L}^{PWF}_{BFT}$ are not only equivalent but identical.

\section{Conclusions}
\label{sec:level7}

The BFT conversion procedure provides a systematics to generate a set of 
first-class theories equivalent to a given second-class one. However, we learnt
that no {\it a priori} statements can be made about the kind of first-class 
theories arising through the BFT mechanism.  

Indeed, in the particle case we
succeded in finding a canonical transformation where all the first-class
constraints $T_a$ became a subset $S_a$ of the transformed momenta.
Then, only the variables $U^a$, canonically conjugate to $S_a$, were affected 
by the gauge transformations. Moreover, it turned out possible to
write the Hamiltonian $K$ in terms gauge independent variables only.
$Q^a$, and $P_a$.

The situation for the SD model was qualitatively different. 
As indicated in Eq.(\ref{524}), only the converted primary constraints 
${\cal T}_0$ and ${\cal T}_i$ turned, after the canonical transformation, into
momenta. The converted secondary constraint ${\cal T}_3$ remains a combination
of the phase-space variables $A^i$ and $P_i$ (see Eq.(\ref{mlett:c524})). As 
consequence, the coordinates of the MCS theory are gauge dependent objects. 

Any attempt of relating the results summarized above with the existence or not
of secondary second-class constraints was destroyed by the Proca-Wentzel 
theory. There, the presence of secondary second-class constraint was not 
enough to prevent us of writing the converted first-class theory only in terms
of gauge invariant fields.  

It has recently appeared in the literature the BFT embeddings of the massive 
Yang-Mills theory\cite{Ban1} and of the non-Abelian SD model\cite{Rothe4}. The
generalization for these cases of our technique mounted on canonical 
transformations is currently under progress.

\end{document}